# Modelling the Effects of User Learning on Forced Innovation Diffusion

Tao Zhang, Peer-Olaf Siebers, Uwe Aickelin

Intelligent Modelling & Analysis Group,

School of Computer Science, University of Nottingham

Tao.Zhang@nottingham.ac.uk

## Abstract

Technology adoption theories assume that users' acceptance of an innovative technology is on a voluntary basis. However, sometimes users are force to accept an innovation. In this case users have to learn what it is useful for and how to use it. This learning process will enable users to transit from zero knowledge about the innovation to making the best use of it. So far the effects of user learning on technology adoption have received little research attention. In this paper - for the first time - we investigate the effects of user learning on forced innovation adoption by using an agent-based simulation approach using the case of forced smart metering deployments in the city of Leeds.

## 1. Introduction

Innovation diffusion theories focus on understanding how, why and at what rate innovative ideas and technologies spread in a social system (Rogers, 1962). In innovation diffusion processes the decisions of whether to adopt the innovation can either be made by the actual users freely and implemented voluntarily, or be made by few authoritative individuals and implemented enforcedly.

Whilst the former type of innovation diffusion received intensive studies (e.g. Griliches, 1957; Mansfield, 1961; Rosenberg, 1972; Geroski, 2000; Hall & Khan, 2002), the latter seems to be an area in its infancy stage. Forced innovation diffusion usually takes place at the level of a massive system or infrastructure upgrade. An example for such a case is a university wide systematic upgrade of the office and lab computer operating system from Windows XP to Windows 7. In this case, the decision is made by the management of the university, and the actual users (e.g. faculty staff and students) are forced to use the innovation with little or even no knowledge about it beforehand and no influence on the choice.

An innovation cannot benefit the society unless its users effectively use it. Thus when forced innovation diffusion happens, it is significantly important to understand how users start to learn the innovative technology, use the technology, and finally make the best use of it and perhaps motivate other users to use it or to improve their knowledge quickly. Users' transition from zero knowledge about an innovation to making the best use of it is a learning process. Understanding the effects of this learning process would help decision-makers design strategies to accelerate users' transition and maximise the benefits that the innovation can bring to the society.

Traditionally, there are some theories aiming to understanding how users (or consumers) learn and adopt an innovation, for example, the Technology Adoption Model (TAM) (Davis, 1989) in information system research and consumer learning models (Solomon et al, 1999) in consumer research. Many of these theories/models are qualitative, static, and only apply to adoption decisions that are made on voluntary basis. In other words, users/consumers seek information and learn knowledge about the product/innovation on their own, and make purchase decisions voluntarily.

Currently no studies extend their application to forced technology adoptions. In this paper, we bridge this academic gap: we draw on the ideas from traditional technology adoption and consumer learning theories/models, and extend the application of them to the forced adoption decisions by using Agent-Based Simulation (ABS).

The agent-based model we have developed is based on the case of smart metering deployments in the UK city of Leeds. This case provides a good example of forced innovation diffusion: the city council uses smart metering energy intervention to systematically upgrade the energy infrastructure in the city, and some energy users are forced to install and use smart meters. With the simulation model we would like to visualise the dynamic process of user learning and understand effects of the learning process on making the best use of the innovation (e.g. effective electricity demand side management).

This paper serves two purposes. On the one hand we want to advance academic knowledge by studying for the first time the field of forced innovation diffusion by extending the application of traditional technology adoption and consumer learning theories to that area via a computational simulation method. On the other hand we aim to provide hands on advice to city council decision-makers with policy implications on how to effective facilitate user learning and maximise the benefits of smart meters for the city.

The paper is structure as follows. In the second section we review relevant theories about consumer learning and technology adoption. In the third section we describe the case study and the developed simulation model and its individual components in detail. We have then conducted a validation experiment. In the fourth section we show the results of this experiment. The fifth section includes our conclusions and an outlook into the future of this project.

## 2. Theoretical Background

Amongst the researchers studying learning processes there is no consensus about how learning happens. Thus the definition of learning is diverse. In psychology, researchers learning as a relatively permanent change in behaviour as a result of increasing experience (Solomon et al, 1999). In marketing, consumer learning is defined as "a process by which individuals acquire the purchase and consumption knowledge and experience they apply to future related behaviour (Schiffman et al, 2008, p. 185). In the real world, individuals learn both directly and indirectly. For example, they can learn from the events that directly influence them; or they can learn from other people's experiences indirectly; sometimes they even learn unconsciously.

Learning covers activities ranging from consumers' responses to external stimuli to a complex set of cognitive processes. There are many learning theories which generally fall into to two categories: behavioural learning and cognitive learning.

### 2.1 Behavioural Learning

Behavioural learning approach takes the assumption that learning happens as a result of responses to external stimuli (Solomon et al, 1999). Thus sometimes behavioural learning theories are also known as stimulus-response theories, as these theories primarily focus on the inputs and outputs in the learning process. The behavioural approach takes the view that a learner's mind is a "black" box, and emphasizes the observable perspectives of behaviour, as shown in Figure 1.

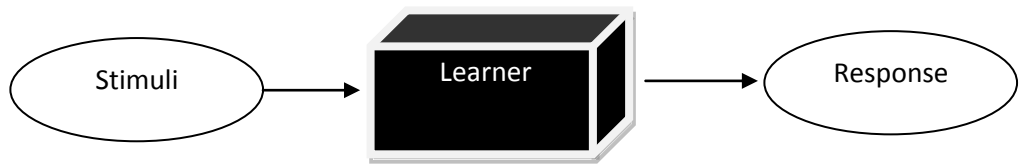

Figure 1: The learner as a "black box" in behavioural learning (Solomon et al, 1999)

In behavioural learning theories learners are mindless passive objects, i.e. they do not make decisions; they could just be taught certain behaviour through repetition or conditioning (Schiffman et al, 2008).

A quantitative expression of the behavioural approach was developed in early marketing literature, e.g. Estes (1950, cited in Bennett and Mandell, 1969), Estes and Buke (1953, cited Bennett and Mandell, 1969), and Bush and Mosteller (1955, cited in Bennett and Mandell, 1969). In all cases learning is treated as a stochastic process and thus response tendencies are treated in probabilistic terms. Howard (1963, cited in Bennett and Mandell, 1969) proposes the following consumer's brand choice learning function:

$$P_A = M(1 - e^{-kt}) \quad (1)$$

Where $P_A$ is the probability of purchasing Brand A; $M$ is the maximum attainable loyalty to Brand A; $k$ is the constant that expresses the learning rate; $t$ is the number of reinforced trials.

This quantitative model (see Figure 2) was empirically validated in Bennett and Mandell (1969) by using the case of new automobile purchase.

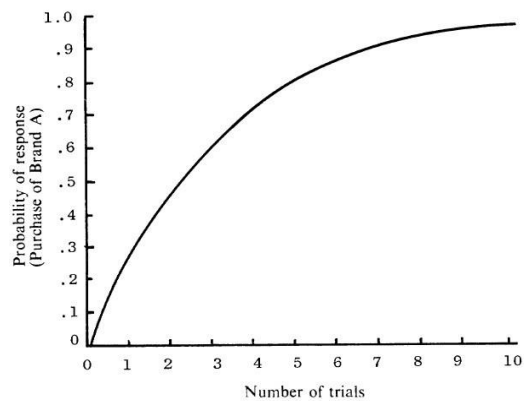

Figure 2: The learning curve for Brand A (Source: Bennett and Mandell, 1969)

## 2.2 Cognitive Learning

The cognitive learning approach assumes that learning is a set of mental processes. In contrast to the behavioural learning approach described earlier the cognitive learning approach takes the view that learners are problem-solvers rather than "black boxes". In other words, learners make purchase/adoption decisions on their own rather than passively repeat trial behaviour. They actively seek information about a product/innovation, process the information and gain motivation or intention to buy/adopt the product/innovation. A typical cognitive learning theory is observational learning theory, which believes that "individuals observe the actions of others and note the reinforcement they receive for their behaviours" (Solomon et al, 1999, p.70). The entire process of observational learning is presented in Figure 3.

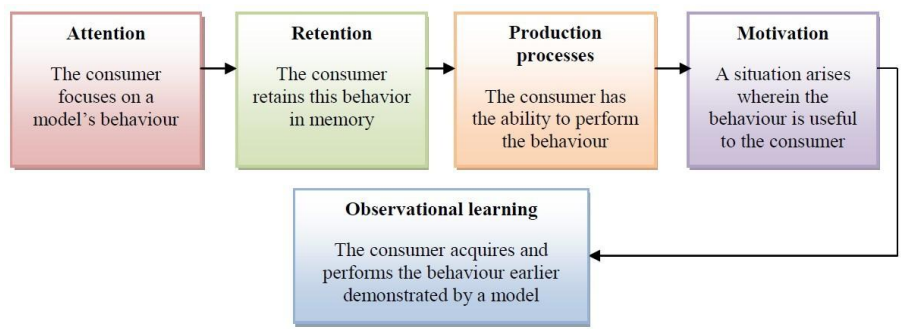

Figure 3: Process of observational learning (Source: Solomon et al, 1999)

**2.3 Our Decision**

Through a review of the two learning approaches we can see that, whilst both approaches are useful and have been empirically validated, a main distinction between the two is whether users/consumers are passive recipients or active decision-makers. Since the study in this paper focuses on forced innovation diffusion in which users passively receive the innovation and are forced to use, we draw on the ideas of behavioural learning approach to develop the simulation model.

**3. The Model**

**3.1 The Case: Smart Metering Energy Interventions in Leeds**

As climate change has become a very important global issue, the UK central government has set a national target: cutting $CO_2$ emission by 34% of 1990 levels by 2020. In the UK there are nearly 70 cities and it is increasingly being suggested that each city will have an important role to play for the country's future sustainability (Keirstead and Schulz, 2010; Bale et al., 2011). With a population of 787,700 (Office for National Statistics, 2011), Leeds City Council (Leeds CC) is the second largest metropolitan council in England and also the UK's largest centre for business, legal and financial services outside London.

Having realised its important role for the future UK sustainability, Leeds CC made a clear statement to voluntarily undertake a target of cutting its $CO_2$ emission by 40% by 2020. However, like many other UK city councils, Leeds CC has no explicit local energy policies/interventions for achieving this target so far. Currently Leeds CC is working with the Universities of Leeds and Nottingham and the research councils on a City Energy Future project, that aims to seek knowledge, experience and develop decision support tools to aid local energy intervention design. Several local energy interventions have been proposed, including setting up a city level strategic council, developing local district heating networks, running energy saving education campaigns and deploying smart meters.

The smart metering intervention has been proposed for two reasons. First, smart metering has been empirically proved to be an effective means to influence energy users' behaviour and reduce energy consumption. Second, many residents living in council-owned properties in Leeds are considered to be in fuel poverty (fuel poverty statistics estimate the number of households that need to spend more than 10% of their income on fuel to maintain a satisfactory heating regime, as well as meeting their other fuel needs (lighting and appliances, cooking and water heating (UK National Statistics, 2011). As Leeds CC has direct control over the council-owned properties, installing smart meters into the council-owned properties would potentially be an effective way to help the occupants get out of fuel poverty.

**3.2 Modelling Rationale**

The research in the paper targets the forced smart metering adoption in Leeds by using Agent-Based Modelling and Simulation (ABMS). ABS is a computational modelling approach to study Multi-Agent Systems (MAS). An Agent-Based Model (ABM) is composed of individual agents, commonly implemented in software as objects. Agent objects have states and rules of behaviour. Furthermore they often have a memory and the capability to evolve over time. This modelling approach lends itself particularly well for modelling people and their behaviour (Siebers and Aickelin, 2011). Running an agent-based model simply amounts to instantiating an agent population, letting the agents behave and interact, and observing what happens globally (Axtell, 2000). Thus a unique advantage of ABS is that almost all behavioural attributes of agents can be captured and modelled. ABS is widely adopted in studying MAS, particularly those with intelligent human beings (e.g. markets, societies, and organisations). In this particular case, the energy users are behave and interact in a community, which is a complex social system that is well suited to agent-based simulation.

### 3.3 Model Design

We use AnyLogic 6.7.0 (XJ Technology, 2011) to develop the model. We design the model based on real map of the Leeds Metropolitan District Area, as shown in Figure 4.

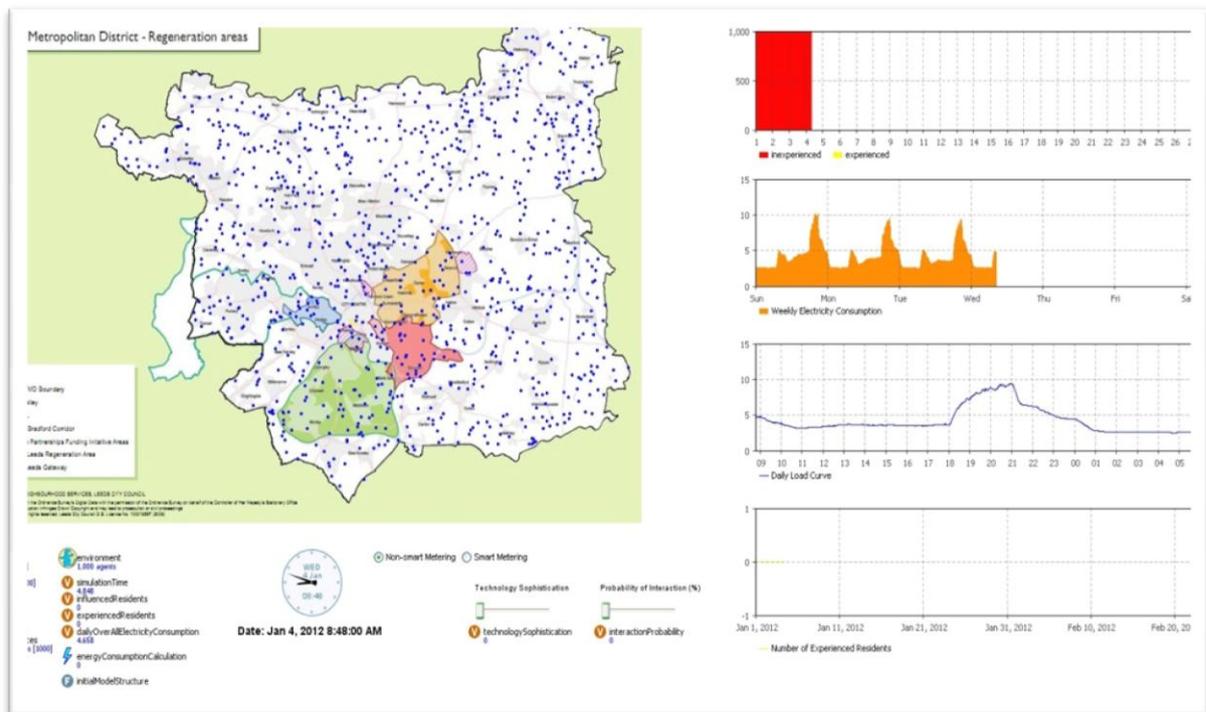

Figure 4: Interface of the model

The blue dots in the map are Residential Energy Consumer (REC) agents. Each REC agent has a set of home electrical appliances. We have carried out a survey to get empirical data about people' social demographic attributes attitude, number and types of electrical appliances, levels of awareness about energy, and life styles in Leeds area (Bale et al., 2011). Based on the empirical data, we have developed some energy consumer archetypes (Zhang et al. 2011). In the model we simulate the daily life of energy consumers. We use a state chart to represent the behaviour of REC agents, as shown in Figure 5.

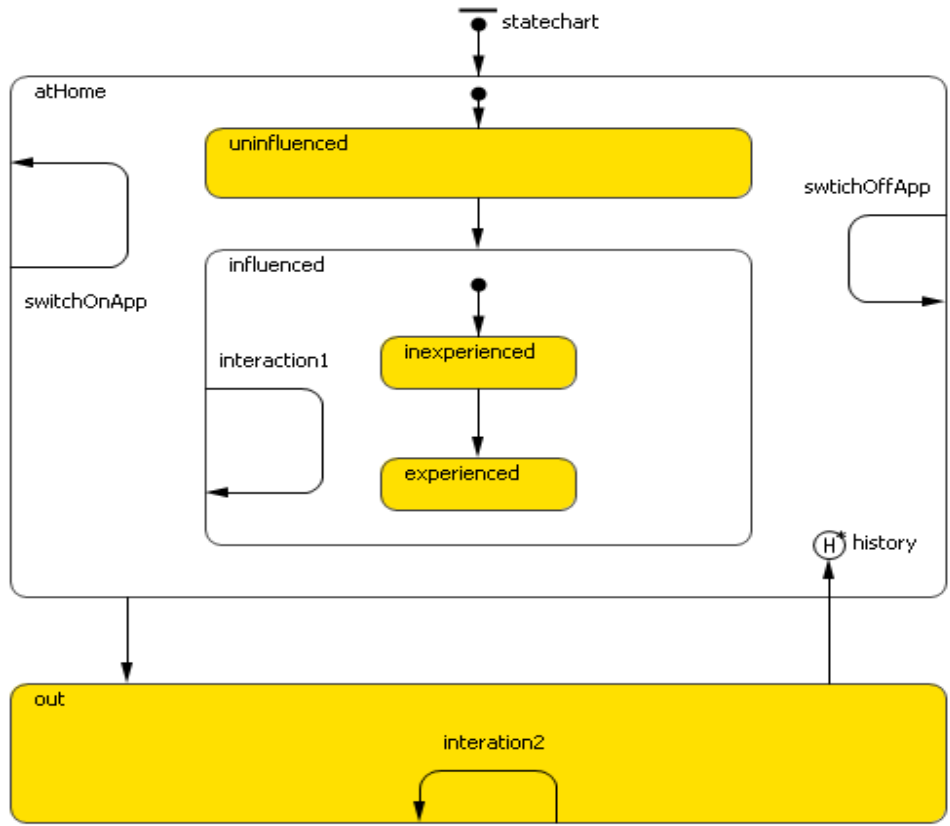

Figure 5: The state chart of REC agents

An REC agent has two main states: "atHome" and "out". The transition from "atHome" state to "out" state is controlled by the parameter "timeLeaveHome", and the transition from "out" state to "atHome" state is controlled by the parameter "timeBackHome". Both parameters are defined by the archetype of the REC agent. For example, a regular work professional often leaves home for work at a time between 8:30am to 9:30am, and goes back home at a time between 5:30pm to 6:30pm.

When the REC agent is in the "atHome" state, it initially is in an "uninfluenced" sub-state if the city council does not implement the smart metering intervention. Otherwise the REC agent transits from the "uninfluenced" sub-state to the "influence" sub-state. When an REC agent is influenced by the city council's smart metering intervention it initially is in the "inexperienced" sub-sub-state. Then the REC agent starts the learning process by trying to use the smart meter. Based on the learning theory chosen in Section 2.3, the more it tries the technology, the more experience it gains about the technology. The transition from the "inexperienced" sub-sub-state to the "experienced" sub-sub-state is a probability threshold ($P_{th}$), which in the simulation model is set at 0.85. The probability ($P_t$) for the REC agent's transition from the "inexperienced" sub-sub-state to the "experienced" sub-sub-state is calculated according to the learning function (see eq. 1). If $P_t$ is larger or equal to $P_{th}$, the REC agent makes the transition.

When an REC agent is at home, it can switch on/off appliances. The behaviour is reflected by the two internal transitions "switchOffApp" and "switchOnApp". The probabilities for them to occur are determined by the level of experience of the individual REC agent.

When an REC agent is influenced by the smart metering energy intervention, it can interact with other REC agents through a network to exchange the knowledge and experience. This is reflected by the two internal transitions "interaction1" and interaction2". In the model, we have chosen a small world network to model the communication channels between REC agents and the probabilities for them to interact with other REC agents are decided by their levels of energy saving awareness.

## 4. Experimentation

In order to conduct some initial tests with our model we created 1000 REC agents in a virtual community and carried out two initial experiments. The first experiment is meant for model validation. We run the model and gained the daily load curve of the virtual community during winter time as shown in Figure 6a. We then compared the result with the real domestic load curve provided by the National Grid as shown in Figure 6b (winter curve). The comparison shows that the two patterns of domestic load curves are quite similar to each other and improves our confidence in the validity of the simulation model.

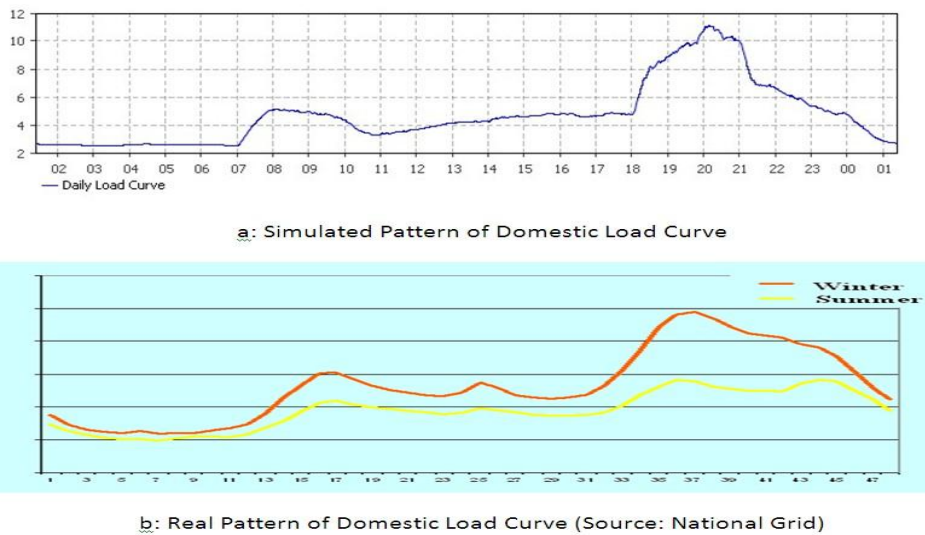

Figure 6: Simulated Load Curve vs. Real Load Curve

In the second experiments, we simulated the REC agents' transition from the "inexperienced" to the "experienced" sub-sub-state. The purpose of the experiment is to show the effects of user learning on the community's energy consumption. We run the model, and gain two domestic load curves in both, the "inexperience" scenario (Figure 7a, all REC agents are inexperienced) and the "experienced" scenario (Figure 7b, 90% of the REC agents are inexperienced). The comparison of these figures shows that the REC agent's transit from "inexperienced" to "experienced" can cause substantial reduction of energy consumption at peak times.

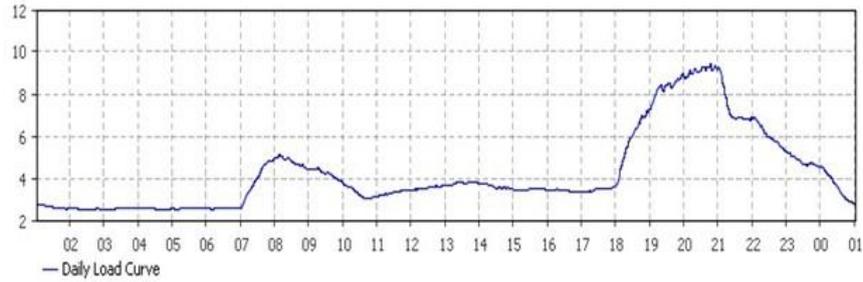

a: inexperienced (all REC agents are inexperienced)

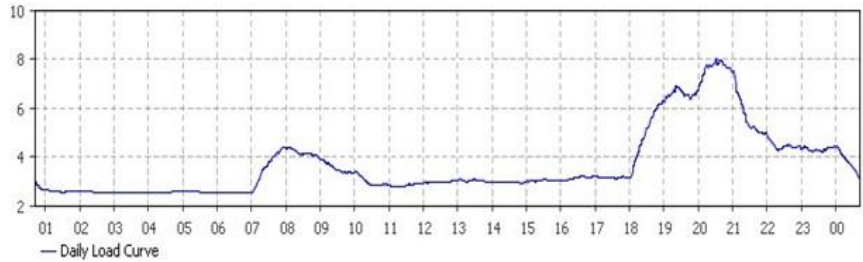

b: experienced (90% of the REC agents are experienced)

Figure 7: "inexperienced" vs. "experienced" REC agents

## 5. Conclusions

In this paper, an agent-based model for studying the effects of user learning on forced smart metering technology adoption is described. This model was developed based on the case of smart metering energy intervention in Leeds. The initial simulation results show that (1) the model can reproduce real world pattern of domestic load curve and (2) that users' learning can enable them to transit from inexperienced in smart metering to experienced with smart metering, which can substantially reduce the energy consumption at the peak time.

With the model, we can carry out further experiments to investigate the effects of user learning on forced technology adoption. For example, we can investigate whether the level of sophistication of the technology can influence user learning; whether different types of social network can facilitate user learning, etc. Thus the model has a huge potential for further studies.

Our study for the first time targeted forced technology adoption, an area that has been ignored in in most other studies. It also provides a first draft of a decision support tool that can be of great use for city council decision makers to support the decisions required to reach the ambiguous UK targets of cutting $CO_2$ emission by 34% of 1990 levels by 2020. Local councils could use this tool to design local energy policies/interventions (adapted to the local needs) to facilitate user learning and maximise the benefits of smart metering.

Our plan is to extend the model in two directions. First we want to consider how forced users can influence others who can voluntarily decide whether to adopt the technology. A second interesting behaviour we would like to consider is counter-intuitive behaviour. Some users use the technology and become experienced but after a certain period they find the technology is not as good as they thought in the first place and finally give it up. We would like to see how this behaviour can be demonstrated in our model, and investigate the effects of the behaviour on technology adoption.